# The effects of aggregation and protein corona on the cellular internalization of iron oxide nanoparticles

M. Safi, J. Courtois, M. Seigneuret, H. Conjeaud and J.-F. Berret*

*Matière et Systèmes Complexes, UMR 7057 CNRS Université Denis Diderot Paris-VII, Bâtiment Condorcet
10 rue Alice Domon et Léonie Duquet, 75205 Paris (France)*

**Abstract :** Engineered inorganic nanoparticles are essential components in the development of nanotechnologies. For applications in nanomedicine, particles need to be functionalized to ensure a good dispersibility in biological fluids. In many cases however, functionalization is not sufficient : the particles become either coated by a corona of serum proteins or precipitate out of the solvent. In the present paper, we show that by changing the coating of iron oxide nanoparticles from a low-molecular weight ligand (citrate ions) to small carboxylated polymers (poly(acrylic acid)), the colloidal stability of the dispersion is improved and the adsorption/internalization of iron towards living mammalian cells is profoundly affected. Citrate-coated particles are shown to destabilize in all fetal-calf-serum based physiological conditions tested, whereas the polymer coated particles exhibit an outstanding dispersibility as well as a structure devoid of protein corona. The interactions between nanoparticles and human lymphoblastoid cells are investigated by transmission electron microscopy and flow cytometry. Two types of nanoparticle/cell interactions are underlined. Iron oxides are found either adsorbed on the cellular membranes, or internalized into membrane-bound endocytosis compartments. For the precipitating citrate-coated particles, the kinetics of interactions reveal a massive and rapid adsorption of iron oxide on the cell surfaces. The quantification of the partition between adsorbed and internalized iron was performed from the cytometry data. The results highlight the importance of resilient adsorbed nanomaterials at the cytoplasmic membrane.

# I – Introduction

Nanotechnology´s ability to shape matter at the scale of the nanometer has opened the door to new generations of diagnostics, therapeutics, imaging agents and drugs for detecting and treating a number of physiological disorders [1, 2]. Engineered inorganic nanoparticles (NPs) are one of the key actors of this strategy. NPs are mostly made of metallic or rare-earth atoms with size-related optical or magnetic properties. These objects are promising agents since their sizes approach those of biological entities such as viruses, proteins or DNA [3]. Several issues become however critical when the NPs are used *in vivo*, including colloidal stability, opsonization by the circulating immunoglobulins, clearance by the immune system and in some cases loss of physical properties. To circumvent these limitations, particles are usually coated with ligands or polymers, or functionalized with peptide sequences [2, 4]. These coating and functionalization strategies are a prerequisite for an accurate delivery to target cells and organs. Owing to the intrinsic complexity of biological fluids however, these efforts are not always sufficient.

On the physico-chemical side, it has been realized recently that the interactions between living cells and NPs depend dramatically on the behavior of the NPs in biological fluids. Two main scenarios have been reported in the recent literature. When dispersed in biological fluids, the NPs can either *i)* become coated by a corona made of serum proteins and of other biomacromolecules, or *ii)* aggregate in a kinetically driven process and form clusters of various sizes. In *i)*, with NPs surrounded by proteins, the corona will constitute the primary contact to the cells [5-11]. Recent reports described this phenomenon as ubiquitous and independent of the type of nanoparticles investigated [11]. Organic particles such as carboxylated polystyrene, and inorganic NPs including nanometer-sized FePt, gold and quantum dots were shown to display the corona effect. The kinetics of adsorption and the resilience of the adsorbates on the inorganic cores were also investigated in details [9] and revealed both equilibrium and non-equilibrium processes, yielding in some cases irreversible adsorption [6, 8].

In case *ii)*, the NPs destabilize and aggregate into micron-size clusters. The destabilization and clustering of NPs in biological fluids have been reported repeatedly during the last decade and appear as a rather frequent phenomenon [12-18]. In such conditions, "what the cell will see" is again not inorganic cores but rather large chunks of particles with eventually proteins and other biomacromolecules on the outer surface. The hydrodynamic properties of aggregated nanomaterials, including diffusion, sedimentation and coupling to a flow are also drastically changed with respect to single NPs, and so are the interactions with cells. A well-known instance is that of sedimentation of micron-size clusters where, due to their own weight NP aggregates enter in contact with the cell membrane of adherent cells [19, 20]. Although a general description is still missing, it is believed that the colloidal destabilization of NPs in biological fluids results from the adsorption of biological molecules (as in the protein corona adsorption), but also from the exchange and/or the removal of the protecting adlayer. As a result, the original coating is not sufficient to offset the van der Waals attraction, which leads to aggregation. Destabilization processes depend on many parameters, in particular on the nature and robustness of the coated adlayer [12, 16, 18, 20].

In the present paper, we investigate the interactions between anionically charged iron oxide (maghemite, $\gamma$-$Fe_2O_3$) NPs and living mammellian cells. The coating of the magnetic 8.3 nm cores was ensured by low molecular weight ligands (citric acid) and by ion-containing polymers (poly(acrylic acid)), both being carboxylate terminated. In a recent report [20], interactions between these NPs and adherent murine fibroblasts were studied, and showed the dramatic effect of the coating on the internalization. From this survey, it was concluded that the strong uptake of the citrate-coated particles was related to the destabilization of the dispersion in the physiological conditions and their sedimentation down to the cell membranes. To avoid the sedimentation bias, we focus here on human lymphoblastoid cells growing in suspension. The particles were first investigated with respect to their stability in biological fluids. The mechanisms of precipitation of the citrate coated particles in the cell culture medium with and without serum were identified at both microscopic and thermodynamic levels. By contrast, the polymer coated NPs were stable in the same conditions and did not display the protein corona effect discussed previously. The interactions between NPs and the human lymphoblastoid cells were also studied by a combination of transmission electron microscopy (TEM) and flow cytometry [21, 22]. We found that within the first minutes of contact, the citrate-coated NPs are massively adsorbed on the cell membranes and are still there after 24 h of incubation. Only a portion of the adsorbed NPs enter the cytoplasm via membrane-bound endosomal compartments, as revealed by TEM. A quantification of the partition between adsorbed and internalized iron was realized from the cytometry data, highlighting the importance of adsorption in NP/cell interactions. By contrast, the polymer coated NPs interacted only marginally with the cells [20, 23].

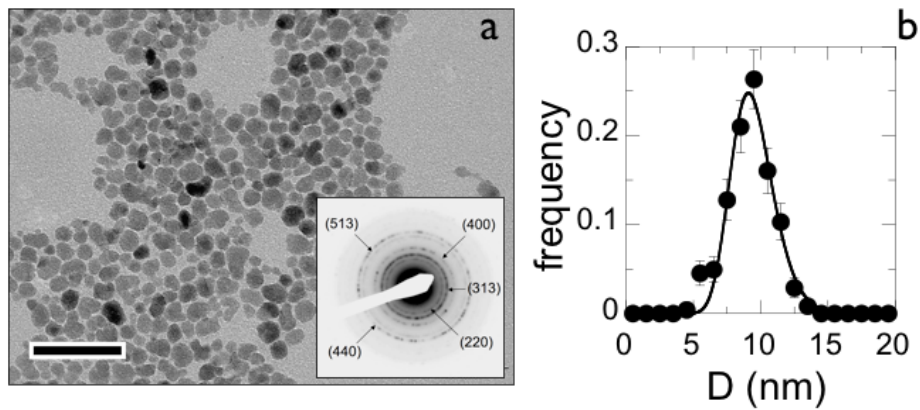

***Figure 1*** : *a) Transmission electron microscopy of iron oxide γ-Fe₂O₃ at the magnification of ×120000. The bar is 40 nm. Inset : Bragg scattering rings obtained by electron beam microdiffraction and identifying the structure of maghemite (Supporting Information SI-1). b) Size distribution of the γ-Fe₂O₃ nanoparticles adjusted by a log-normal distribution function (continuous line) with median diameter 8.3 nm and polydispersity 0.26.*

# II – Materials and Methods
## II.1 – Chemicals, synthesis and characterization
*Nanoparticles synthesis* : The iron oxide nanoparticles (bulk mass density $\rho$ = 5100 kg m$^{-3}$) were synthesized according to the Massart method [24] by alkaline co-precipitation of iron(II) and iron(III) salts and oxidation of the magnetite ($Fe_3O_4$) into maghemite ($\gamma$-$Fe_2O_3$). The nanoparticles were then size-sorted by subsequent phase separations [24, 25]. At pH 1.8, the particles were positively charged, with nitrate counterions adsorbed on their surfaces. The resulting particle-particle electrostatic repulsion imparted an excellent colloidal stability of the dispersions, typically over several years. An image of transmission electron microscopy (TEM) obtained from a dilute aqueous dispersion exhibits compact and spherical particles (Fig. 1a). The size distribution of the particles was described by a log-normal function with a median diameter 8.3 nm and polydispersity 0.26 (Fig. 1b). The polydispersity was defined as the ratio between standard deviation and average diameter. The inset in Fig. 1a illustrates the electron microdiffraction scattering obtained for the $\gamma$-Fe$_2$O$_3$ particles. The crystallinity of the nanoparticles was demonstrated by the appearance of five diffraction rings which wave vectors matched precisely those of the maghemite structure (Supporting Information SI-1). In contrast to recent investigations [26], in the pH range 1.8-10 and for different coating and environments probed, the release rate of ferric ions was found to be extremely low (Supporting Information SI-2). In this work, the NP concentration are defined by the percentage by weight of $\gamma$-Fe$_2$O$_3$ in the dispersion or by the iron molar concentration [Fe]. With these units, c = $8\times10^{-2}$ wt. % corresponds to [Fe] = 10 mM.

*Coating* : To improve their colloidal stability, different types of coating based on the electrostatic adsorption of an organic adlayer on the particles were developed. The complexation of the surface charges with ligands such as citrate ions was performed during the particle synthesis. It allowed to reverse the surface charge of the particles from cationic at low pH to anionic at high pH, through a ionization of the carboxyl groups. At pH 8, the particles were stabilized by electrostatic interactions mediated by the anionically charged ligands. As a ligand, citrate ions were characterized by adsorption isotherms, *i.e.* the adsorbed species were in equilibrium with free citrates molecules dispersed in the bulk. For the citrate coated particles (Cit–$\gamma$-Fe$_2$O$_3$), the structural charge density was ascertained at -2$e$ nm$^{-2}$ by conductivity [27, 28] and light scattering measurements [29], resulting in average structural charges of -430$e$. The affinity constant between the citrate ligands and the maghemite surface was derived from the adsorption isotherm and estimated at 1560 M$^{-1}$ (Supporting Information SI-3).

The cationic particles were also coated with M$_W$ = 2000 g mol$^{-1}$ poly(acrylic acid) (PAA$_{2K}$) using the *precipitation-redispersion* process [30]. The polymer of polydisperstiy 1.7 was purchased from Sigma Aldrich and used without further purification. To adsorb the PAA$_{2K}$ chains onto the particle surface, the particles were first precipitated into a dilute solution containing a large excess of polymers (pH 1.8). The precipitate was separated by centrifugation and its pH was increased by addition of ammonium hydroxide. The precipitate redispersed spontaneously at pH 7 - 8, yielding a clear solution that contained the individual polymer coated particles, dubbed PAA$_{2K}$–$\gamma$-Fe$_2$O$_3$ in the following. This process resulted in the adsorption of a highly resilient 3 nm polymer adlayer surrounding the particles. The thickness layer was later determined by dynamical light scattering. The density of chargeable carboxylic groups was evaluated by acid titration at 25 $\pm$ 3 nm$^{-2}$. For the particles of diameter 8.3 nm, it corresponded to 5400 structural anionic charges in average. As a final step, the dispersions were all dialyzed against DI-water which pH was first adjusted to 8 (Spectra Por 2 dialysis membrane with MWCO 12 kD). At this pH, 90 % of the carboxylate groups of the citrate and PAA$_{2K}$ coating were ionized.

## II.2 – Experimental Methods

*Transmission Electron Microscopy (TEM)* : Electron microscopy on nanomaterials was carried out on a Jeol-100 CX microscope at the SIARE facility of Université Pierre et Marie Curie (Paris 6). It was utilized for the characterization of the uncoated $\gamma$-Fe$_2$O$_3$ particles using magnifications ranging from 10000$\times$ to 160000$\times$. An example of TEM image is illustrated in Fig. 1. For TEM experiments on cells, the 2139 lymphoblasts were first seeded onto a 6-well plate and incubated with the citrate and PAA$_{2K}$-coated particles during 10 min, 1 h and 24 h. Cells were then fixed with 2.5% glutaraldehyde in 0.1 M sodium cacodylate buffer, at pH 7.4 and room temperature for 2 hours. Fixed cells were washed in 0.2 M PBS. After washing, samples were post-fixed for 1 hour in a 1% osmium-phosphate buffer 0.1 M for 45 min at room temperature in dark conditions. The samples were then dehydrated in reagent-grade ethanol and embedded in an araldite CY212 epoxy resin. 90 nm ultra-thin sections were cut with an ultra-microtome (LEICA, EM UC6), stained with uranyl acetate and lead citrate, and then examined with a TECNAI-12 electron microscope (Philips).

*Light scattering* : Static and dynamic light scattering were performed on a Brookhaven spectrometer (BI-9000AT autocorrelator, $\lambda$ = 632.8 nm) for measurements of the scattered intensity $I_S$ and of the diffusion constant $D_{dif}$. In this study, the suspending solvents were 18.2 M$\Omega$ de-ionized water, phosphate buffer (PBS) or RPMI cell culture medium with or without fetal calf serum (FCS). The concentrations targeted in the present work were in the dilute regime, and comprised between [Fe] = 0.01 mM and 10 mM, corresponding to $8\times10^{-5}$ - $8\times10^{-2}$ wt. %. Using dynamical light scattering, the collective diffusion coefficient $D_{Diff}$ was determined from the second-order autocorrelation function of the scattered light. From the value of the coefficient, the hydrodynamic diameter of the colloids was calculated according to the Stokes-Einstein relation, $D_H = k_BT/3\pi\eta_SD_{Diff}$, where $k_B$ is the Boltzmann constant, T the temperature (T = 298 K) and $\eta_0$ the solvent viscosity. The viscosity of the various solvents used in this work were measured as a function of the temperature between 15 and 40 °C (Supporting Information SI-4). The autocorrelation functions were interpreted using the cumulant and the CONTIN fitting procedure provided by the instrument software.

In the dilute regime of concentration, the scattering intensity varies as $I_S$ scales with the product $KM_Wc$, where K is the scattering contrast, $M_W$ and c the molecular weight and weight concentration of the scatterers. To evaluate the NP stability in the solvents mentioned previously, the scattering intensity and hydrodynamic diameter were monitored as a function of the time. If the dispersion destabilizes, the particles aggregate and both $I_S$ and $D_H$ increase due to the augmentation of the size and molecular weight of the elementary scatterers. If the dispersion remains stable, $I_S$ and $D_H$ are flat and equal their initial values.

In some flow cytometry experiments, the suspending medium turned out to be turbid immediately upon addition of iron oxide. The turbidity was interpreted as a destabilization and an aggregation of the NPs. To study the kinetics of aggregation, light scattering experiment was performed using the same protocol as for incubation. A test tube containing 1 ml of the cellular medium (without cells) was placed on the spectrometer and a small volume (~ 10 µL) of a concentrated iron oxide dispersion was poured rapidly in the tube and the sample was homogenized with a pipette. The final NP concentration was [Fe] = 1 mM, corresponding to c = $8\times10^{-3}$ wt. %. The scattering intensity $I_S$ and the hydrodynamic diameter $D_H$ were then monitored as a function of the time for 2 hours after the NP fast injection, a time that was considered to be sufficient to conclude about the NP behavior in the solvent [23].

*Human lymphoblastoid cells and culture*

Lymphoblast are one of the different stages of physiological differentiation inside the lymphoïd line leading to the lymphocytes. This differentiation stage appears after the activation of the small lymphocytes, inside the lymph nodes by an antigen in the organism. The lymphoblastoid cell line 2139 used in this study is a normal line provided to us by Dr Janet Hall from the Institut Curie (Orsay, France). This cell line was immortalized by the virus Epstein-Barr (EBV) which was obtained by Dr Gilbert Lenoir from the Institut Gustave Roussy (Villejuif, France). These immortalized cells were mostly used as a model or control for different studies in ataxia-telangiectasia-mutated dependent functions like the kinase activity, cell cycle progression and viability [31-33]. 2139 human lymphoblastoid cells were grown in T25-flasks in RPMI with high glucose (2.0 g L$^{-1}$) and stable glutamine (PAA Laboratories GmbH, Austria). This medium was supplemented with 10% fetal bovine serum (FBS) and 1% penicillin/streptomycin (PAA Laboratories GmbH, Austria), referred to as cell culture medium. Exponentially growing cultures were maintained in a humidified atmosphere of 5% $CO_2$ and 95% air at 37°C and passaged twice weekly. The cells were pelleted by centrifugation at 1200 rpm for 5 min. Supernatants were removed and cell pellets were resuspended in assay medium and counted using a Malassez counting chamber.

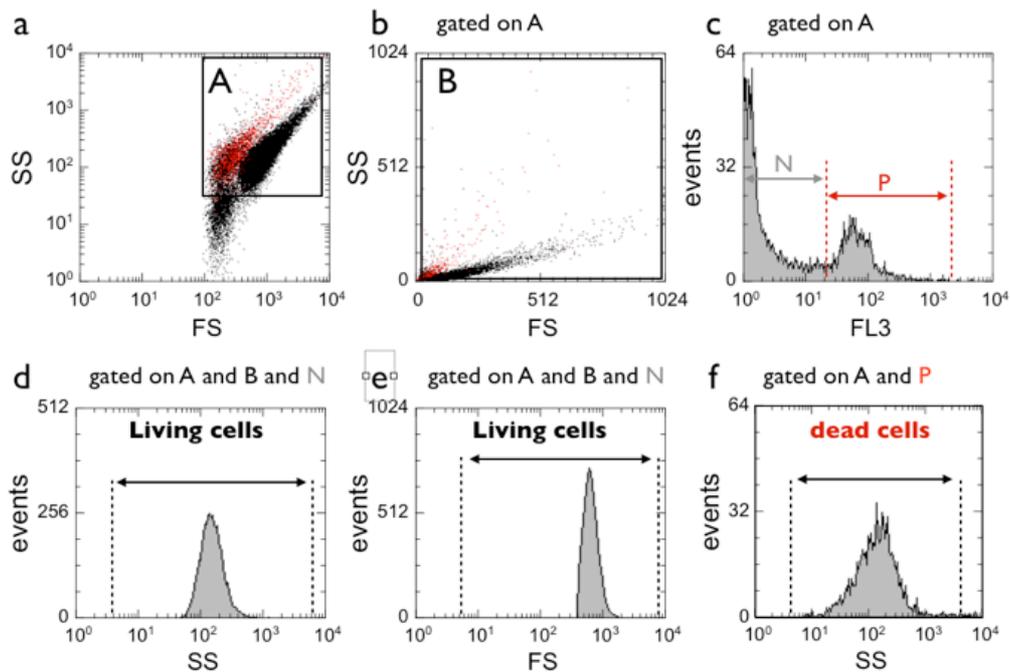

*Figure 2 :* Cytometric analysis: *a)* and *b)* Dot plots of the side scatter (SS) intensities versus forward scatter (FS) intensities in double logarithmic and in linear scale, respectively. In a), Gate A includes the dead and the living cells. In b), Gate B excludes part of the dead cells. *c)* Histogram of the red fluorescence intensities of events in gate A (double logarithmic scale). *d)* Histogram of the side scatter intensities of events in gates A and B and N corresponding to living cells. *e)* Same as in d) but for the forward scatter intensities. *f)* Histogram of the side scatter intensities of events in gates A and P for the dead cells.

*Flow Cytometry* : Forward scatter (FS), side scatters (SS) and propidium iodide (PI) fluorescence of individual cells incubated for various times with nanoparticles were measured on an EPICS 4C flow cytometer (Beckman Coulter). Forward scatter FS and SS intensities (measured at 488 nm) were digitalized on both linear and logarithmic scale (4 decades). Red fluorescence intensities emitted by dead cells (measured at 610 nm) was digitalized on a logarithmic scale. Cellular debris and most of the NP aggregates were eliminated by using a large gate in a FS *versus* SS cytogram (Figs. 2a and 2b, gate A). Dead cells and living cells were identified by their forward scatter and red fluorescence intensities (gates N and P in Fig. 2c). FS and SS histograms (logarithmic scale) of living cells (A and B and N, Figs. 2d and 2e) and dead cells (gate A and P, Fig. 2f) were visualized and median values calculated in each experimental condition. Nanoparticles at concentrations ranging from 0.3 to 10 mM of iron were mixed with exponentially growing 2139 cells ($10^6$ ml$^{-1}$) in complete medium supplemented with 1% HEPES and incubated for various times (indicated in the legends), at 4°C or 37°C, under constant agitation to prevent sedimentation of cells and particle aggregates. Two minutes before cytometric analysis, propidium iodide (10 μM final concentration) was added to the cell suspension.

# III – Results and discussion
## III.1 – Colloidal stability in physiological media
### III.1.1. – Stability of the citrate coated nanoparticles in RPMI
Figs. 3a and 3b display the scattering intensity ($I_S$) and hydrodynamic diameter ($D_H$) *versus* time for 5 different solvents, which were PBS1x, RPMI, RPMI containing citrate ions at concentration 1 mM and 10 mM, and DI-water containing calcium nitrate (Ca(NO$_3$)$_2$) and magnesium chloride (MgCl$_2$). The molar concentrations of Ca$^{2+}$ and Mg$^{2+}$ were identical to those present in the RPMI formulation (Supporting Information SI-5). In PBS1x, the Cit–γ-Fe$_2$O$_3$ NPs are found to be stable, as both the intensity and the diameter remain unchanged. In RPMI by contrast, both $I_S$ and $D_H$ increase drastically and display characteristic S-shaped curves as a function of the time. Starting at 19 nm, $D_H$ reaches rapidly 300 nm after 2 min, and build up to values of about 3 μm. With a progressive addition of citrate ions in RPMI, from 0.1 mM to 10 mM, the initial increase is delayed and it eventually ceases for citrate concentration above 5 mM. The data obtained in the presence of 10 mM of citrates demonstrate that the Cit–γ-Fe$_2$O$_3$ dispersion with an excess of ligands is again stable for more than two hours, a result in good agreement with the literature [34, 35]. The rapid aggregation kinetics observed with RPMI, as well as the stabilizing effect of the citrate ligands above 5 mM suggest that in RPMI the ligands desorb from the NP surfaces because they are preferentially complexed by chemical species present in the RPMI formulation. To get a better insight into this mechanism, all compounds present in RPMI at concentrations larger than 50 mg L$^{-1}$ were tested against the citrate-coated particles. These chemicals are inorganic salts, such as calcium nitrate, magnesium sulfate, sodium chloride etc…, amino acids such as L-alanyl-L-glutamine, L-glutamine, or other components such as D-glucose. The list of these compounds is provided in Supporting Information, Section SI-5. Among this list, only calcium and magnesium ions at concentrations equal or above those of the RPMI formulation gave positive results. Upon addition of Ca$^{2+}$ and Mg$^{2+}$ ions, the destabilization kinetics exhibits the same S-shaped curve as that of pure RPMI (Fig. 3a). Isothermal titration calorimetry experiments (data not shown) performed by titrating calcium chloride with tri-sodium citrate ions yielded an affinity constant of $1.5 \times 10^4$ M$^{-1}$, *i.e.* 10 times larger than that of citrate for the maghemite surface (Supporting Information SI-3). These findings indicate that the divalent calcium and magnesium cations present in the cell medium are causing the precipitation of Cit–γ-Fe$_2$O$_3$ NPs through a preferential complexation with the ligands.

### III.1.2. – Stability of the citrate coated nanoparticles in complete cell culture medium
In a second set of experiments, the role of the serum on the stability of Cit–γ-Fe$_2$O$_3$ NPs in RPMI was outlined. Figs. 3c and 3d show the intensities and diameters for three media : RPMI containing 10 vol. % of fetal bovine serum, RPMI with FBS and 10 mM of citrate ions and a cellular medium composed of RPMI and 10 mM citrate that was incubated with lymphoblasts during 24 h. This later medium was considered because we anticipated that the cell released species might be active towards the NP surfaces. In the three cases, a similar behavior is observed : both $I_S$ and $D_H$ exhibit an initial and rapid increase. $D_H$ for instance jumps abruptly from 20 to 40 – 50 nm within the first seconds, after which it grows slowly up to 100 nm. After a few hours, the particles sedimented at the bottom of the test tube. Interestingly, this initial jump of 20 - 30 nm corresponds to twice the size of the proteins/biological molecules present in the complete culture medium (Supporting Information SI-6). These findings are interpreted as the formation of a protein corona around the particles at short times, and as a destabilization of the dispersion at long times. Here the kinetics is slower than that of Figs. 3a and 3b, and does not impart a S-shape profile. From the above results, it can be concluded that in the conditions of the cell cultures with 10 vol.% FBS and at iron concentrations relevant for *in vitro* [5, 12, 16, 18, 21, 36-38] and *in vivo* [38, 39] assays, Cit–γ-Fe$_2$O$_3$ NPs are colloidally unstable, and precipitate.

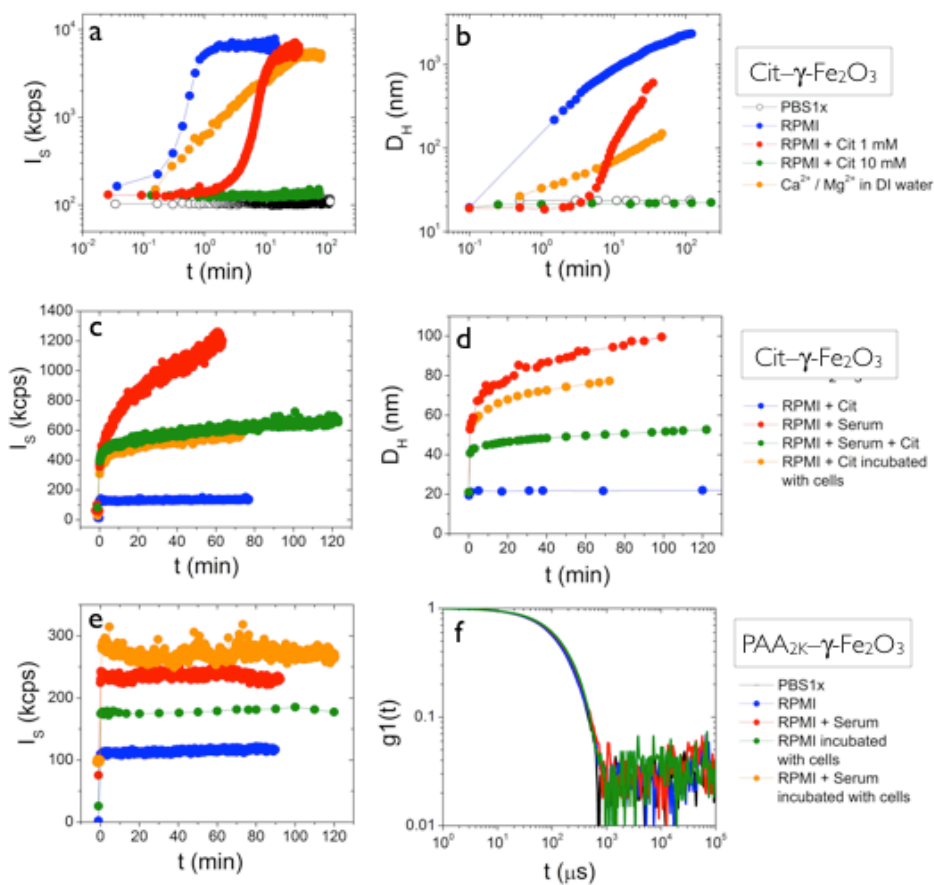

**Figure 3 : a)** and **b)** *Scattering intensity ($I_S$) and hydrodynamic diameter ($D_H$) as a function of the time for Cit–γ-Fe$_2$O$_3$ NPs dispersed in PBS1x, RPMI, RPMI containing citrate ions at concentration 1 mM and 10 mM, and DI-water containing calcium nitrate and magnesium chloride. The S-shaped curves found for $I_S$ and $D_H$ are the signatures of a rapid destabilization of the dispersion.* **c)** and **d)** *$I_S$ and $D_H$ versus time for Cit–γ-Fe$_2$O$_3$ NPs dispersed in RPMI containing 10 vol. % of fetal bovine serum, RPMI with FBS and 10 mM of citrate ions and a cellular medium composed of RPMI and 10 mM citrate that was incubated with lymphoblasts during 24 h. The data are interpreted at short times as the formation of a protein corona and at long times later as a destabilization of the dispersion.* **e)** *Scattering intensity versus time for the same solvent as in a) for PAA$_{2K}$–γ-Fe$_2$O$_3$ NPs.* **f)** *First order autocorrelation function $g^{(1)}(t)$ for PAA$_{2K}$–γ-Fe$_2$O$_3$ particles at [Fe] = 25 mM in PBS1x, RPMI, RPMI with serum and RPMI with serum incubated with the lymphoblasts during 24 h.*

### III.1.3. – Stability of the PAA$_{2K}$ coated nanoparticles in complete cell culture medium

In a third experiment, the same iron oxide cores, coated this time with PAA$_{2K}$ were investigated by light scattering in various conditions. As for the Cit–γ-Fe$_2$O$_3$, the media were RPMI, RPMI containing 10 vol. % FBS, and a cellular medium composed of RPMI (with or without FBS) that was incubated with the lymphoblasts for 24 h. Fig. 3e displays the intensities *versus* time after the fast injection of the $PAA_{2K}-\gamma-Fe_2O_3$ NPs in the media. There, $I_S$ and $D_H$ (data not shown) remain flat for two hours, indicating that the dispersions are indeed stable. These results confirm those obtained with the same particles and coating using Dulbecco's Modified Eagle's Medium (DMEM) as culture medium [20, 23]. Light scattering is a very accurate technique which can detect an increase in diameter of a few nanometers [40]. Fig. 3f shows the first order autocorrelation function $g^{(1)}(t)$ for particles at [Fe] = 25 mM ($c_{\gamma-Fe2O3} = 2 \times 10^{-2}$ wt. %) in 4 media : PBS1x, RPMI, RPMI with serum and RPMI with serum incubated with the lymphoblasts during 24 h. In the four assays, the autocorrelation functions are found to superimpose, indicating an identical distribution of the NPs sizes. Derived from the second coefficient of the cumulant analysis, these diameters are respectively $D_H$ (± 1 nm) = 28.4, 28.6, 27.8 and 28.8 nm with a polydispersity of 0.2. From these data, it can be concluded that $PAA_{2K}$ coated particles are colloidally stable in RPMI-based culture media on a broad range of concentrations, and are devoid of a protein corona. Because the binding affinity of $PAA_{2K}$ towards the NP surfaces is high, the polymeric segments are not removed by the modifications of the physico-chemical conditions. As such, the $PAA_{2K}-\gamma-Fe_2O_3$ should be regarded as a kinetically frozen organic/inorganic hybrid rather than a NP characterized by dynamical and reversible exchanges of its surface ligands. As shown below, the propensity of the particles to remain disperse or to aggregate play a noticeable role on the interactions with living cells.

## III.2 –Transmission Electron Microscopy

The lymphoblasts seeded with iron oxide NPs were further investigated by TEM. Figs. 4 and 5 provide representative images from lymphoblastoid cells incubated with $Cit-\gamma-Fe_2O_3$ and $PAA_{2K}-\gamma-Fe_2O_3$, respectively. The experimental conditions were an incubation time of 10 min (Fig. 4c and Fig. 5a) and 24 h (in Fig. 4b, 4d and Fig. 5b) and a NP concentration of 0 mM (control, Fig. 4a), 1 mM (Fig. 4b) and 10 mM (Fig. 4c, 4d, 5a and 5b). A careful analysis of the TEM images allows us to identify two distinctive locations for the NPs, either adsorbed at the surface of the cells, or in the cytoplasm in membrane-bound compartments. Among the 100 images of cells collected for this study, NPs were never found in the cytosol nor in the nucleus. Figs. 4c and 4d compare the impact of the incubation time on the uptake of $Cit-\gamma-Fe_2O_3$ at [Fe] = 10 mM. In both images, the adsorbed NPs form an heterogeneous 200 – 500 nm thick layer. In the field of view of Fig. 4, the layer surrounds almost entirely the cell. Close-up views of the delimited areas (c1) and (d2) indicate that the NPs are actually aggregated into polydisperse clusters, either directly deposited on the plasma membrane or linked to it through other clusters. At 10 min, the topology of the coated membrane is rough and irregular, whereas it is much smoother at 24 h. The close-view (c1) in Fig. 4 highlights an invagination where the NP-coated membrane has the morphology of a nascent endosome. In the cytoplasm, membrane-bound compartments are also visible and have an average diameter of 500 nm, such as in the close-up views (c1) and (d1).

Figs. 4b and 4d illustrate the dose effect on the NP/cell interactions and provide a comparison of the data obtained at [Fe] = 1 mM and [Fe] = 10 mM. The same patterns as those discussed previously are observed. NP clusters are adsorbed onto the membrane (b1) or enclosed into endosomal compartments (b2). At 1 mM however, both adsorbed and internalized clusters are less and also smaller than those found at 10 mM, an observation that was later confirmed by light scattering and by flow cytometry. The close-up views (b1) and (b2) reveal an average size in the 100 nm range.

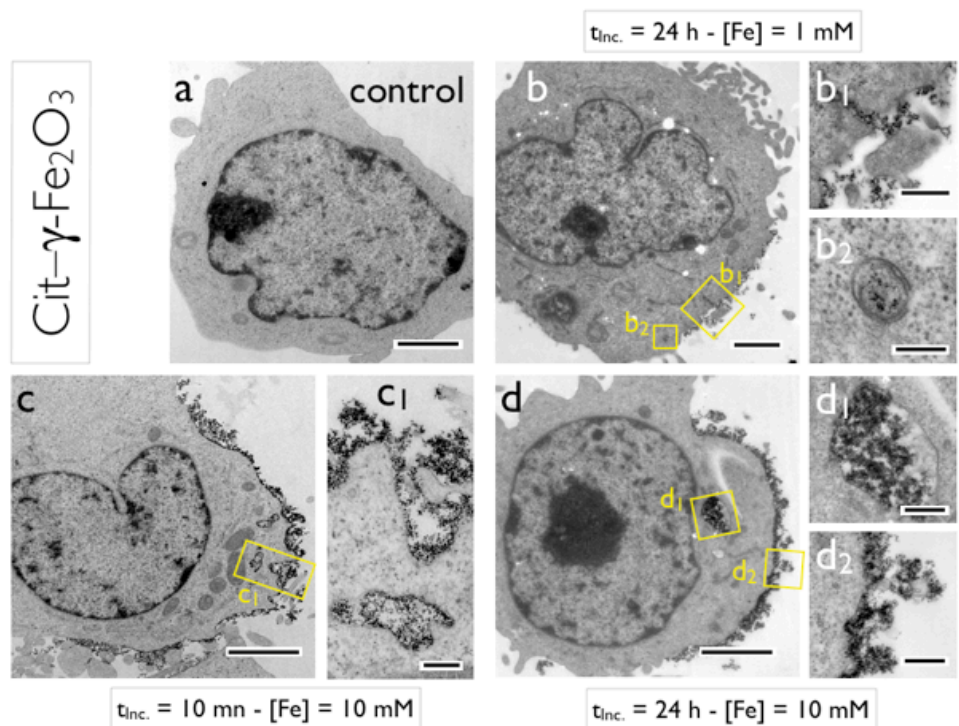

**Figure 4 :** *Transmission electron microscopy (TEM) images of lymphoblastoid cells incubated with Cit–γ-Fe₂O₃ NPs.* ***a)*** *Control.* ***b)*** *Cells incubated for 24 h at [Fe] = 1 mM; b1) and b2) are close views of the delimited areas in b).* ***c)*** *Cells incubated for 10 min at [Fe] = 10 mM; c1) displays a close-up view of the cell surface.* ***d)*** *cells incubated for 24 h at [Fe] = 10 mM; d1) and d2) show NPs enclosed in an endosome and NP clusters adsorbed on the cell membrane, respectively. Bars in a-d) are 2 μm, 200 nm in b2) and c1), 300 nm in d1) and d2) and 500 nm in b1).*

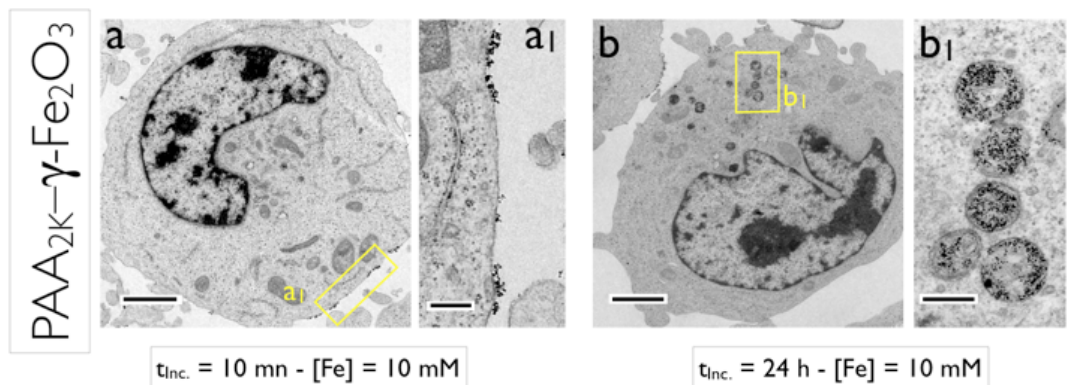

**Figure 5:** *Transmission electron microscopy (TEM) images of lymphoblastoid cells incubated with PAA₂ₖ–γ-Fe₂O₃ NPs.* ***a)*** *Cells incubated for 10 min at [Fe] = 10 mM; a1) is a close view of the cell membrane delimited in a).* ***b)*** *Cells incubated for 24 h at [Fe] = 10 mM; b1) displays NPs enclosed in 200 nm endosomes. Bars in a-b) are 2 μm, 400 nm in a1) and 300 nm in b1).*

Figs. 5a and 5b depict the representative behavior of PAA₂ₖ–γ-Fe₂O₃ NPs after incubation times of 10 min and 24 h, respectively ([Fe] = 10 mM). In marked contrast with the citrate coated NPs, very few particles can be visualized on these images. The NPs are again located either at the surface (a1) or in endosomes (b1). At the surface, the particles are single or clustered into aggregates of a few units. These clusters have an average size of 50 nm or less. After 24 h of incubation, the NPs are removed from the cell surface and localized in membrane-bound compartments of average size of 250 nm. Within these compartments (b1) the particles are randomly spread and not aggregated, again a noticeable difference with Cit–γ-Fe₂O₃. Put together, these findings lead to the conclusions that although the particles have the same magnetic core and are carrying the same carboxylate groups at their surfaces, the internalization rates and amounts of Cit–γ-Fe₂O₃ and PAA₂ₖ–γ-Fe₂O₃ NPs are very different. Similar results were reported for the NIH/3T3 adherent fibroblasts [20].

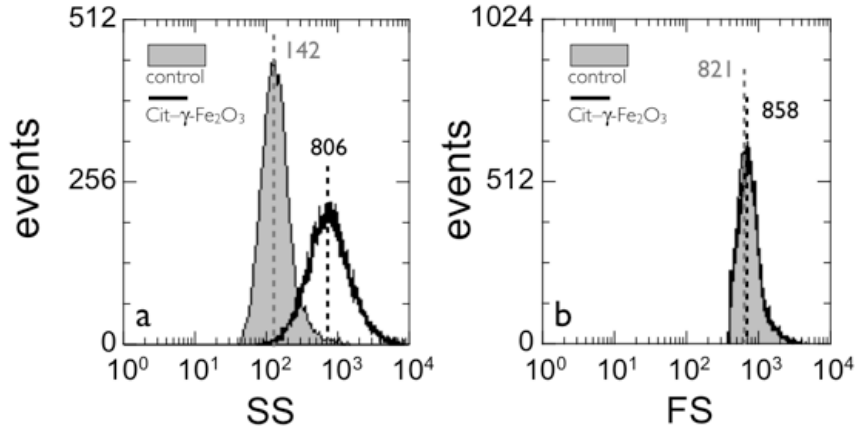

***Figure 6 : a)*** *Side Scatter (SS) intensity of living cells incubated for 4 hours with, or without, Cit–γ-Fe₂O₃ coated NPs at [Fe] = 10 mM.* ***b)*** *Same as in a) but for the Forward Scatter (FS) intensity. The intensity distributions were fitted by a log-normal distribution function shown in Eq. 1. For the SS distributions, the median values (shown by dotted lines) increased with [Fe] but the polydispersity remained constant. The SS intensity changes due to the adsorption/internalization of NPs by cells were expressed as ratio $R_{SS}$ of the median intensities of cells incubated with NPs over the mean values of control cells. The FS distributions were mostly unchanged upon NPs addition.*

## III.3 – Flow cytometry

### III.3.1 – Comparison between citrate and PAA2K coated particles

In flow cytometry experiments, the forward (FS) and side (SS) scatter intensities generated by the cell illumination are probing the cellular size and refractive index respectively. Figs. 6a and 6b show respectively the FS and SS intensities of the 2139 lymphoblastoid cells that were incubated with Cit–γ-Fe₂O₃ NPs. The experimental conditions were an incubation time of 4 h at 37 °C and a concentration [Fe] = 10 mM. Under these conditions, the SS intensity exhibits a shift from 142 ± 7 in arbitrary units for the control to 806 ± 40 for the incubated cells, whereas the FS signal remains unchanged at 821 ± 41 and 858 ± 43 respectively. This result indicates that the refractive index of the lymphoblastoid cells was enhanced because of their interactions with the NPs, but that their average sizes did not vary noticeably. Similar results were reported on Chinese hamster ovary cells [22] and on human hepatoma cells treated with titanium dioxide and with starch-coated magnetic NPs [21]. A quantitative analysis of the cytometry spectra revealed that both SS and FS intensities could be well accounted for by a log-normal distribution :

$$p(I, I_0, s) = \frac{1}{\sqrt{2\pi}\beta(s)I} \exp\left(-\frac{\ln^2(I / I_0)}{2\beta(s)^2}\right) \tag{1}$$

where I denotes the FS or SS intensities, $I_0$ their median values and $\beta(s)$ is related to the polydispersity s *via* the relationship $\beta(s) = \sqrt{\ln(1+s^2)}$. The polydispersity is defined as the ratio between the standard deviation and the average intensity. For the data collected in this study, the polydispersity s is found to be $0.50 \pm 0.05$ whatever the incubation time and concentration. As a result, the cytometry data were characterized by a unique parameter, noted $R_{SS}$. $R_{SS}$ is defined as the median side scatter intensity obtained from the seeded cells, divided by the same quantity obtained from the control cells : $R_{SS} = I_{0,NP}^{SS} / I_{0,control}^{SS}$. With these notations, $R_{FS} = I_{0,NP}^{FS} / I_{0,control}^{FS} = 1$, as shown in Fig. 6b. In the sequel of the paper, the $R_{SS}$-ratios are investigated as a function of the incubation time $t_{Inc}$ and of the iron concentration [Fe].

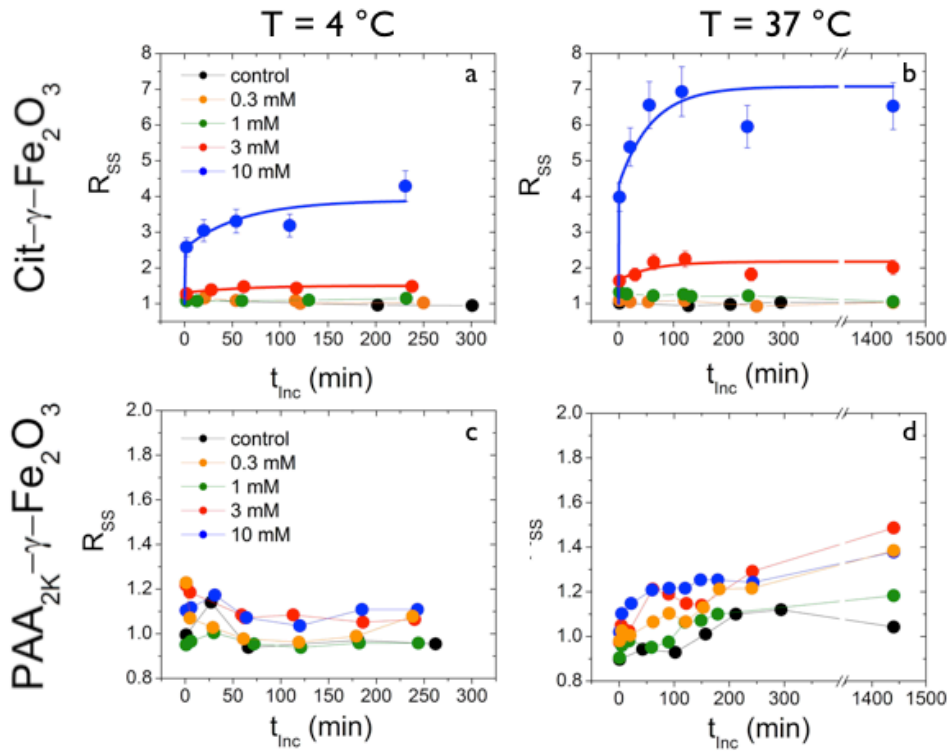

**Figure 7 :** *Transient behaviors of the cellular side scatter intensity (SS) induced upon incubation of human lymphoblastoid cells with Cit–γ-Fe₂O₃ and PAA₂ₖ–γ-Fe₂O₃ NPs. The investigated concentrations are [Fe] = 0 (control), 0.3, 1, 3 and 10 mM at temperatures T = 4° C (a,c) and T = 37° C (b,d). The results are expressed as the ratios $R_{SS}$ between the median SS intensity obtained from seeded cells, divided by the same quantity obtained from the control (Eq. 1).*

Fig. 7 displays the variations of the $R_{SS}$-ratios with the incubation time at 4 °C (a,c) and 37 °C (b,d) upon addition of citrate (a,b) and of PAA₂ₖ (c,d) coated particles. The data are compared to those of untreated cells. The investigated concentrations were [Fe] = 0.3, 1, 3 and 10 mM. As noted in earlier reports [21, 22, 26], the evolution of the side scatter intensities recorded at 37 °C imparts two interaction mechanisms of the NPs : their adsorption on the plasma membrane and their internalization into endocytic vesicles. The later mechanism is blocked at 4 °C and so the $R_{SS}$-ratios observed at this temperature reflect only the adsorption of the NPs at the cell surface.

Upon addition of NPs at 0.3 mM and 1 mM, the $R_{SS}$-ratios do not increase significantly. For $PAA_{2K}-\gamma-Fe_2O_3$, weak increases are also observed at 3 and 10 mM at the two temperatures ($R_{SS}$ = 1.4 at 10 mM, see Fig. 7d). In contrast, the addition of $Cit-\gamma-Fe_2O_3$ NPs at 3 mM and 10 mM induces strong $R_{SS}$ increases : within the first minutes of contact with NPs (first data point in Fig. 7a and 7b), the $R_{SS}$-ratio jumps to an initial value noted $R_{SS}(t_{Inc} \rightarrow \infty)$ which depends on both concentration and temperature. The initial jump is followed by a saturation (noted $R_{SS}(t_{Inc} \rightarrow \infty)$) after 2 hours for all concentrations tested. This initial rise corresponds to about 2/3 and 1/2 of the stationary 4°C and 37°C values. The continuous lines between the data points are exponential growth functions of the form :

$$R_{SS}(t_{Inc}) = R_{SS}(t_{Inc} \rightarrow 0) + \Delta R_{SS}\left(1 - \exp(-\frac{t_{Inc}}{\tau})\right) \qquad (2)$$

where $\Delta R_{SS} = R_{SS}(t_{Inc} \rightarrow \infty) - R_{SS}(t_{Inc} \rightarrow 0)$ denotes the shift between the initial and stationary states and $\tau$ the characteristic time of the kinetics. Note that the $\Delta R_{SS}$'s at 4 °C are weaker than at 37 °C, indicating that during the transient phase the NPs were indeed internalized. For the four profiles, one finds $\tau$ = 60 ± 10 min. Note also that for $PAA_{2K}$-coated NPs there is no initial jump of the side scatter ratio, i.e. $R_{SS}(t_{Inc} \rightarrow 0)$ = 1.

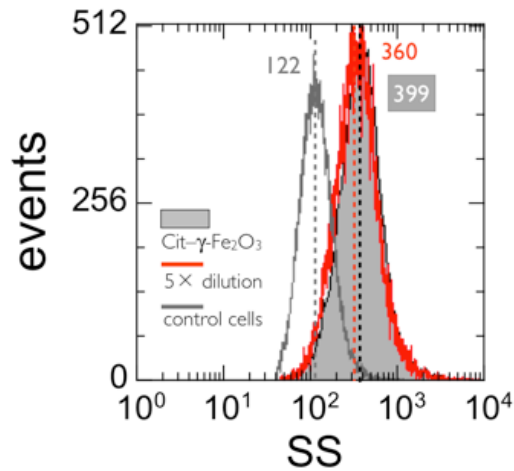

**Figure 8** : *Side scatter intensity histograms of IP negative cells incubated at 4 °C for 60 minutes with 10 mM Cit–γ-Fe₂O₃ NPs (full grey), of the same suspension diluted 5 times and incubated 10 min more at 4 °C (red line), and of control cells (grey line). Numbers over the histograms indicated median values of the distributions.*

### III.3.2 – Effect of dilution

To ensure that SS intensities were not contaminated by particles passing trough the laser beam at the same time as the cells, we compared the intensity distributions obtained from two samples, one for which the cells were incubated with $Cit-\gamma-Fe_2O_3$ ($t_{Inc}$ = 1 h, [Fe] = 10 mM and T = 4 °C) and a second that was diluted five times with RPMI. Working here at 4 °C guarantees that the NPs were not internalized. As shown in Fig. 8, the SS intensities are peaked at 399 ± 20 and 360 ± 18, and have similar $R_{SS}$ (3.27 and 2.95 respectively). This assay demonstrates that the side scatter increases resulted only from interactions of NPs with lymphoblastoid cells, and not from NP aggregates of comparable sizes. Furthermore, the analysis of the same sample at different times after the dilution (30 min, 1 and 2 h, data not shown) reveals no significant change in the intensity distributions, with $R_{SS}$-values similar to those of the undiluted sample. These findings underline the remarkable stability of the NP/cell interactions. The previous tests were repeated under different $t_{Inc}$ and [Fe] conditions and the results were similar. In conclusion of this part, we infer that Cit–$\gamma$-Fe$_2$O$_3$ aggregates heavily adsorb onto the cell surfaces, forming in some cases layers up to 500 nm. The adsorbed are not prone to desorb after 4 h.

*III.3.3 – Concentration dependence of the side scatter intensity*

As seen in Figs. 7, the stationary values of the side scatter $R_{SS}(t_{Inc} \to \infty)$ depend on iron oxide dose in a nonlinear manner. For citrate coated particles at 4 °C, *i.e.* in absence of active internalization processes, it is around 1 at 1 mM and reaches a value of 4 at 10 mM ($t_{Inc} = 240$ min). A possibility to account for the concentration dependence of the $R_{SS}$-ratios is to assume that cells interact mostly with aggregates of particles, and that the sizes of these aggregates depend on the concentration. To support this assumption, light scattering was performed on Cit–$\gamma$-Fe$_2$O$_3$ NPs dispersed in RPMI at different concentrations between 0.01 and 10 mM. After 4 hours, the hydrodynamic diameters were measured and found to grow from 70 nm at [Fe] = 0.01 mM to 700 nm at [Fe] = 10 mM (Fig. 9). Studied by electrophoresis using Zetasizer Nano ZS Malvern Instrument, the clusters were found to be electrostatically neutral, with a $\zeta$–potential round $0 \pm 5$ mV. Under physiological conditions, citrate coated particles precipitate and form neutral clusters, which spontaneously adsorb on the cellular membranes when incubated with cells.

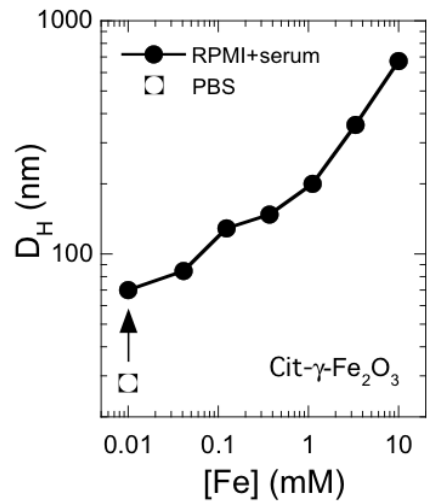

**Figure 9 :** *Hydrodynamic diameter ($D_H$) of Cit–$\gamma$-Fe$_2$O$_3$ NPs dispersed in a complete cell culture medium (RPMI + 10 vol. % serum) as a function of the iron concentration [Fe]. The incubation was set to 4 h. The diameters were found to be larger than to that of particles dispersed in PBS1x and to increase with time and concentration*

*III.3.4 – Quantitative determination of the adsorbed and internalized iron*

To further assess the role of the aggregation state on the cellular refractive index, we took advantage of the presence of the 2 - 5% of dead cells that were present in all the cell samples investigated. Dead cells were distinguished by their strong red fluorescence linked to the nuclear incorporation of propidium iodide present in the suspension medium. With such permeant cells, NP concentrations should quickly equilibrate between the cell interior and exterior and the measure of the SS intensity changes should discriminate between two hypotheses :

    1.   The small variations of the SS intensity for the $PAA_{2K}-\gamma$-$Fe_2O_3$ NPs (Fig. 7c and 7d) are due to low levels of adsorbed/internalized materials.

    2.   Because these particles remained unaggregated in the culture medium, their contribution to the cellular refractive index is weak, and so are the SS intensity shifts.

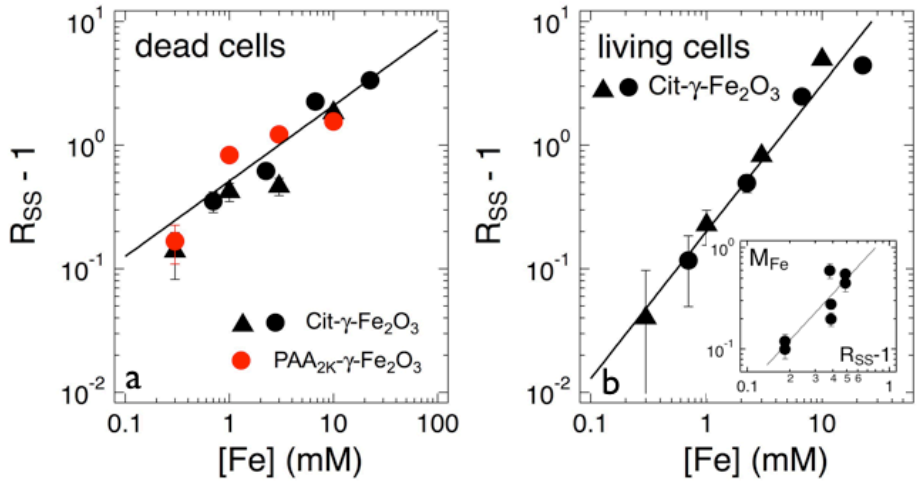

**Figure 10 : a)** *Concentration dependence of the quantity $R_{SS}-1$ measured for dead lymphoblasts incubated with $Cit-\gamma$-$Fe_2O_3$ and $PAA_{2K}-\gamma$-$Fe_2O_3$ during 5 min at $T = 4°C$. Dead cells were identified by their red fluorescence linked to the nuclear incorporation of propidium iodide. The straight line shows a power law variation with an exponent of 0.6.* **b)** *Same as in Fig. 10a, but for living cells incubated with $Cit-\gamma$-$Fe_2O_3$ for $t_{Inc} = 4$ h at $T = 37$ °C. Together with the scaling proposed in the inset, the straight line of equation* $R_{SS} - 1 = 0.2[Fe]^{1.2}$ *was later used to translate the side scatter intensities into mass of iron interacting with the cells.* **Inset :** *Experimental relationship between the mass of iron internalized and the quantity $R_{SS}$-1, as obtained from cells seeded with $PAA_{2K}-\gamma$-$Fe_2O_3$.*

Fig. 10a shows the quantity $R_{SS}-1$ as a function of [Fe] for $Cit-\gamma$-$Fe_2O_3$ and $PAA_{2K}-\gamma$-$Fe_2O_3$ NPs for the dead 2139 lymphoblasts ($t_{Inc} = 5$ min; $T = 4°C$). The data points were found to be well fitted by a power law with exponent 0.6 (straight line in Fig. 10a). The figure evidences that the presence of citrate or $PAA_{2K}$-coated particles inside permeant cells induces similar increases of their SS intensities, regardless of their state of aggregation. It also indicates that the cellular refractive index is correlated to the total amount of iron present inside the cells, and not to its spatial distribution.

Fig. 10b illustrates the concentration dependence of the quantity $R_{SS}-1$ measured for living cells incubated with $Cit-\gamma$-$Fe_2O_3$ ($t_{Inc} = 4$ h and $T = 37$ °C). The figure discloses again a power behavior of the form $R_{SS} - 1 = 0.2[Fe]^{1.2}$ observed over two decades in concentration. Complementary studies discussed in Supporting Information (SI-7) describe a method to measure the amount of internalized and/or adsorbed iron oxide. This protocol was applied to lymphoblastoid cells seeded with $PAA_{2K}-\gamma$-$Fe_2O_3$ but not with $Cit-\gamma$-$Fe_2O_3$ NPs because of the presence in this later sample of large aggregates that could not be separated from the cells (e.g. by magnetophoresis). Here the $PAA_{2K}-\gamma$-$Fe_2O_3$ particles serve as a calibration of the mass of internalized iron per cell, $M_{Fe}$. The inset in Fig. 10b displays the relationship between the internalized mass of iron and the quantity $R_{SS}-1$ for this compound. Using this calibration curve, the $R_{SS}(t_{Inc})$-evolutions in Fig. 7 were translated into $M_{Fe}(t_{Inc})$. The masses of iron per cell incubated with 10 mM of Cit–γ-Fe$_2$O$_3$ particles at 4 °C and 37 °C are shown in Fig. 11. After 2 min of incubation, $M_{Fe}$-data exhibit a jump from 0 to 2.8 pg/cell at 4 °C and from 0 to 7.4 pg/cell at 37 °C, showing that adsorption at 37 °C is twice that observed at 4 °C. At both temperatures, the initial jump is followed by a levelling-off and a saturation of the masses after two hours at levels reaching 7.0 pg/cell at 4°C and 21.6 pg/cell at 37°C. All together, these results suggest that at 37 °C, two thirds of the particle interacting with cells (~15 pg/cell at 10 mM Fe) are on the cell surfaces, while one third is internalized (7 pg/cell) These values are in good agreement with those of the literature on similar compounds [35, 41, 42].

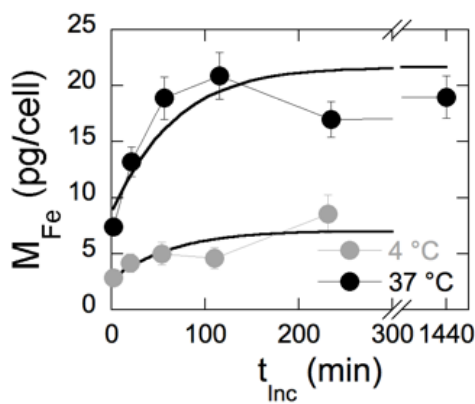

***Figure 11*** *: Mass of iron $M_{Fe}$ interacting with the human lymphoblastoid cells as a function of the incubation time. The data are for Cit–γ-Fe$_2$O$_3$ at [Fe] = 10 mM and the continuous lines are from best fits exponential growth laws similar to Eq. 2.*

# V – Conclusion

In the present paper we have established a relationship between the phase behavior of anionically coated iron oxide NPs in biological fluids and their interactions with mammalian cells growing in suspension. Iron oxide NPs coated with negatively charged citrate ligands are found to destabilize in cell culture media and to interact strongly with human lymphoblastoid cells. Particles made from the same magnetic core, but coated with a 3 nm anionic polymer adlayer exhibit much weaker interactions with cells. These findings confirm those obtained with adherent murine fibroblasts and treated with the same particles [20]. The destabilization kinetics of Cit–γ-Fe$_2$O$_3$ NPs in biological fluids was monitored by static and dynamic light scattering. The precipitation of the Cit–γ-Fe$_2$O$_3$ NPs in biological fluids without serum is attributed to the displacement of the citrates from the particle surfaces towards the bulk, as they are preferentially complexed by calcium and magnesium counterions of the culture medium. In the presence of 10 vol. % serum, the aggregation is slowed down due to the presence of a protein corona surrounding the particles. Given the rather wide use of citrates for coating engineered NPs (as in the cases of gold [8, 43] and rare-earth oxide [29, 44-46]), the present results, together with those of the recent literature [47-49] confirm the rather poor coating performances of such low-molecular weight ligands. By contrast, in all biological fluids tested, the polymer coated NPs exhibited a remarkable colloidal stability and an absence of protein corona. The reasons for the outstanding stabilization behavior of polymer-coated particles were reviewed recently [50]. Polymers at interfaces form a dense brush which acts as a physical barrier against aggregation, even in complex media. In addition, when the polymers are bearing anionic charges as in the present specimen, the organic layer imparts an additional electrostatic repulsion between the tethered surfaces, resulting in an electrosteric interaction [23].

In this study, we ascertain two types of NP-cell behaviors : NPs are either adsorbed on the cellular membranes, or internalized into membrane-bound endocytic compartments. The visualization of these cytoplasmic compartments was made possible thanks to transmission electron microscopy. In the case of adsorption, the layer of materials deposited on the membranes can grow up to 500 nm and remains unperturbed after 24 h of incubation, pointing out the remarkable stability of the NP/cell interactions. Flow cytometry on treated lymphoblastoid cells displays important shifts of the intensity distribution as compared to control cells. We demonstrate here that this effect was due to an increase in the scattering contrast (and not in the size) of the seeded cells passing through the laser beam. The exploitation of flow cytometry data to evaluate the uptake of NPs in mammalian cells was alluded to by Suzuki et al. [22], and later on confirmed by Xia et al. [21]. Our results support this earlier work. Here, the displacements of the intensity distribution are analyzed as a function of the coating, incubation time, dose and temperature. Working both at 4 °C and at 37 °C allowed us to discriminate between the adsorbed and internalized NPs. The kinetics of NP/cell interactions reveal two main features : for the precipitating citrate coated particles, the cytometry signal at 4 °C exhibit a massive and rapid adsorption of iron oxide on the cell surfaces. At this point, it is unclear whether the NPs precipitated directly on the cellular membranes, or if there was first the formation of aggregates and then a deposition of the cells. As anticipated, the polymer coated NPs exhibit a reduced adsorption and/or uptake towards the cells. Extended analysis of the cytometry data emphasizes the possibility to quantify the intensity shifts in terms of internalized and adsorbed amounts, a technique that could find relevant applications in *in vitro* experimentation.

## Acknowledgments

We thank Loudjy Chevry, Jérôme Fresnais, Jean-Pierre Henry, Olivier Sandre, Minhao Yan for fruitful discussions. The Laboratoire Physico-chimie des Electrolytes, Colloïdes et Sciences Analytiques (UMR Université Pierre et Marie Curie-CNRS n° 7612) is acknowledged for providing us with the magnetic nanoparticles and with their characteristics. Berengère Abou and Rémy Colin are acknowledged for their support in the light scattering experiments. We thank Mabel San Roman and Alexis Canette from the imaging facility (ImagoSeine) at the Jacques Monod Institute (Paris, France) for their precious help in the transmission electron microscopy studies. This research was supported in part by the Agence Nationale de la Recherche under the contracts BLAN07-3_206866 and ANR-09-NANO-P200-36, by the European Community through the project : "*NANO3T—Biofunctionalized Metal and Magnetic Nanoparticles for Targeted Tumor Therapy*", project number 214137 (FP7-NMP-2007-SMALL-1) and by the Région Ile-de-France in the DIM framework related to Health, Environnement and Toxicology (SEnT).

## Supporting Information

The Supporting Information section shows the full characterization of the iron oxide nanoparticle in terms of size and crystallinity (SI-1). The release amounts and release rates of ferric ion $Fe^{3+}$ at neutral and acidic pH for the Massart dispersions used in this work are estimated in SI-2. The adsorption isotherm of citrate ions on $\gamma$-$Fe_2O_3$ nanoparticles is analyzed using the Langmuir formulation in SI-3. Static viscosity of the all RPMI biological fluids used in this work are provided in SI-4. The composition of cellular medium RPMI 1640 is recalled in SI-5 whereas light scattering properties of cell culture medium yield the size of the biological molecules and proteins present in the medium (SI-6). Details of the protocol for measuring the mass of metal internalized/adsorbed by living cells (MILC) are provided in SI-7. This information is available free of charge via the Internet at http://pubs.acs.org/.

# References

1.      Ferrari M. Cancer nanotechnology: Opportunities and challenges. Nature Reviews Cancer 2005;5:161 - 171.

2.      Kim J, Piao Y, Hyeon T. Multifunctional nanostructured materials for multimodal imaging, and simultaneous imaging and therapy. Chemical Society Reviews 2009;38:372-390.

3.      Doak SH, Griffiths SM, Manshian B, Singh N, Williams PM, Brown AP, et al. Confounding experimental considerations in nanogenotoxicology. Mutagenesis 2009;24:285-293.

4.      Gupta AK, Gupta M. Synthesis and surface engineering of iron oxide nanoparticles for biomedical applications. Biomaterials 2005;26:3995-4021.

5.      Mahmoudi M, Shokrgozar MA, Sardari S, Moghadam MK, Vali H, Laurent S, et al. Irreversible changes in protein conformation due to interaction with superparamagnetic iron oxide nanoparticles. Nanoscale 2010;3:1127-1138.

6.      Rocker C, Potzl M, Zhang F, Parak WJ, Nienhaus GU. A quantitative fluorescence study of protein monolayer formation on colloidal nanoparticles. Nat Nanotechnol 2009;4:577-580.

7.      Walczyk D, Bombelli FB, Monopoli MP, Lynch I, Dawson KA. What the Cell "Sees" in Bionanoscience. J Am Chem Soc 2010;132:5761-5768.

8.      Casals E, Pfaller T, Duschl A, Oostingh GJ, Puntes V. Time Evolution of the Nanoparticle Protein Corona. ACS Nano 2010;4:3623-3632.

9.      Jiang X, Weise S, Hafner M, Rocker C, Zhang F, Parak WJ, et al. Quantitative analysis of the protein corona on FePt nanoparticles formed by transferrin binding. J R Soc Interface 2010;7:S5-S13.

10.     Mu QX, Li ZW, Li X, Mishra SR, Zhang B, Si ZK, et al. Characterization of Protein Clusters of Diverse Magnetic Nanoparticles and Their Dynamic Interactions with Human Cells. Journal of Physical Chemistry C 2009;113:5390-5395.

11.     Sund J, Alenius H, Vippola M, Savolainen K, Puustinen A. Proteomic Characterization of Engineered Nanomaterial-Protein Interactions in Relation to Surface Reactivity. ACS Nano 2011;5:4300-4309.

12.     Auffan M, Decome L, Rose J, Orsiere T, DeMeo M, Briois V, et al. In Vitro Interactions between DMSA-Coated Maghemite Nanoparticles and Human Fibroblasts: A Physicochemical and Cyto-Genotoxical Study. Environ Sci Technol 2006;40:4367 - 4373.

13.     Boldt K, Bruns OT, Gaponik N, Eychmuller A. Comparative Examination of the Stability of Semiconductor Quantum Dots in Various Biochemical Buffers. The Journal of Physical Chemistry B 2006;110:1959-1963.

14.     Williams D, Ehrman S, Pulliam Holoman T. Evaluation of the microbial growth response to inorganic nanoparticles. Journal of Nanobiotechnology 2006;4:3.

15.     Diaz B, Sanchez-Espinel C, Arruebo M, Faro J, de Miguel E, Magadan S, et al. Assessing Methods for Blood Cell Cytotoxic Responses to Inorganic Nanoparticles and Nanoparticle Aggregates. Small 2008;4:2025-2034.

16.     Petri-Fink A, Steitz B, Finka A, Salaklang J, Hofmann H. Effect of cell media on polymer coated superparamagnetic iron oxide nanoparticles (SPIONs): Colloidal stability, cytotoxicity, and cellular uptake studies. European Journal of Pharmaceutics and Biopharmaceutics 2008;68:129-137.

17.     Allouni ZE, Cimpan MR, H⁻l PJ, Skodvin T, Gjerdet NR. Agglomeration and sedimentation of TiO2 nanoparticles in cell culture medium. Colloids and Surfaces B: Biointerfaces 2009;68:83-87.

18.     Villanueva A, et al. The influence of surface functionalization on the enhanced internalization of magnetic nanoparticles in cancer cells. Nanotechnology 2009;20:115103.

19.     Teeguarden JG, Hinderliter PM, Orr G, Thrall BD, Pounds JG. Particokinetics in vitro: Dosimetry considerations for in vitro nanoparticle toxicology assessments (vol 95, pg 300, 2007). Toxicological Sciences 2007;97:614-614.

20.     Safi M, Sarrouj H, Sandre O, Mignet N, Berret J-F. Interactions between sub-10-nm iron and cerium oxide nanoparticles and 3T3 fibroblasts: the role of the coating and aggregation state. Nanotechnology 2010;21:10.

21.     Xia J, Zhang S, Zhang Y, Ma M, Xu K, Tang M, et al. The Relationship Between Internalization of Magnetic Nanoparticles and Changes of Cellular Optical Scatter Signal. Journal of Nanoscience and Nanotechnology 2008;8:6310-6315.

22.     Suzuki H, Toyooka T, Ibuki Y. Simple and easy method to evaluate uptake potential of nanoparticles in mammalian cells using a flow cytometric light scatter analysis. Environ Sci Technol 2007;41:3018-3024.

23.     Chanteau B, Fresnais J, Berret J-F. Electrosteric Enhanced Stability of Functional Sub-10 nm Cerium and Iron Oxide Particles in Cell Culture Medium. Langmuir 2009;25:9064-9070.

24.     Massart R, Dubois E, Cabuil V, Hasmonay E. Preparation and properties of monodisperse magnetic fluids. J Magn Magn Mat 1995;149:1 - 5.

25.     Bee A, Massart R, Neveu S. Synthesis of very fine maghemite particles. J Magn Magn Mat 1995;149:6 - 9.

26.     Soenen SJH, Himmelreich U, Nuytten N, Pisanic TR, Ferrari A, De Cuyper M. Intracellular Nanoparticle Coating Stability Determines Nanoparticle Diagnostics Efficacy and Cell Functionality. Small 2010;6:2136-2145.

27.     Dubois E, Cabuil V, Boue F, Perzynski R. Structural analogy between aqueous and oily magnetic fluids. J Chem Phys 1999;111:7147 - 7160.

28.     Lucas IT, Durand-Vidal S, Dubois E, Chevalet J, Turq P. Surface Charge Density of Maghemite Nanoparticles:‚Äâ Role of Electrostatics in the Proton Exchange. The Journal of Physical Chemistry C 2007;111:18568-18576.

29.     Berret J-F. Stoichiometry of electrostatic complexes determined by light scattering. Macromolecules 2007;40:4260-4266.

30.     Berret J-F, Sandre O, Mauger A. Size distribution of superparamagnetic particles determined by magnetic sedimentation. Langmuir 2007;23:2993-2999.

31.     Angele S, Romestaing P, Moullan N, Vuillaume M, Chapot B, Friesen M, et al. ATM haplotypes and cellular response to DNA damage: Association with breast cancer risk and clinical radiosensitivity. Cancer Res 2003;63:8717-8725.

32.     Gutierrez-Enriquez S, Hall J. Use of the cytokinesis-block micronucleus assay to measure radiation-induced chromosome damage in lymphoblastoid cell lines. Mutat Res Genet Toxicol Environ Mutagen 2003;535:1-13.

33.     Gutierrez-Enriquez S, Fernet M, Dork T, Bremer M, Lauge A, Stoppa-Lyonnet D, et al. Functional consequences of ATM sequence variants for chromosomal radiosensitivity. Gene Chromosomes Cancer 2004;40:109-119.

34.     Wilhelm C, Cebers A, Bacri JC, Gazeau F. Deformation of intracellular endosomes under a magnetic field. Eur Biophys J Biophys Lett 2003;32:655-660.

35.     Luciani N, Gazeau F, Wilhelm C. Reactivity of the monocyte/macrophage system to superparamagnetic anionic nanoparticles. J Mater Chem 2009;19:6373-6380.

36.     Brunner TJ, Wick P, Manser P, Spohn P, Grass RN, Limbach LK, et al. In vitro cytotoxicity of oxide nanoparticles: Comparison to asbestos, silica, and the effect of particle solubility. Environ Sci Technol 2006;40:4374-4381.

37.     Rad AM, Arbab AS, Iskander ASM, Jiang Q, Soltanian-Zadeh H. Quantification of Superparamagenetic Iron Oxide (SPIO)-Labeled Cells Using MRI. Journal of Magnetic Resonance Imaging 2007;26:366-374.

38.     Shaw SY, Westly EC, Pittet MJ, Subramanian A, Schreiber SL, Weissleder R. Perturbational profiling of nanomaterial biologic activity. Proc Natl Acad Sci U S A 2008;105:7387-7392.

39.     Lutz J-Fo, Stiller S, Hoth A, Kaufner L, Pison U, Cartier R. One-Pot Synthesis of PEGylated Ultrasmall Iron-Oxide Nanoparticles and Their in Vivo Evaluation as Magnetic Resonance Imaging Contrast Agents. Biomacromolecules 2006;7:3132-3138.

40.     Qi L, Sehgal A, Castaing JC, Chapel JP, Fresnais J, Berret J-F, et al. Redispersible hybrid nanopowders: Cerium oxide nanoparticle complexes with phosphonated-PEG oligomers. Acs Nano 2008;2:879-888.

41.     Rad AM, Janic B, Iskander A, Soltanian-Zadeh H, Arbab AS. Measurement of quantity of iron in magnetically labeled cells: comparison among different UV/VIS spectrometric methods. Biotechniques 2007;43:627-+.

42.     Zhang Y, Kohler N, Zhang M. Surface Modiffication of Superparamagnetic Magnetite Nanoparticles and their Intracellular Uptake. Biomaterials 2002;23:1553 - 1561.

43. Kim D, Park S, Lee JH, Jeong YY, Jon S. Antibiofouling polymer-coated gold nanoparticles as a contrast agent for in vivo x-ray computed tomography imaging. J Am Chem Soc 2007;129:7661-7665.

44. Biggs S, Scales PJ, Leong Y-K, Healy TW. Effects of Citrate Adsorption on the Interactions between Zirconia Surfaces. J Chem Soc Faraday Trans 1995;91:2921 - 2928.

45. Munnier E, Cohen-Jonathan S, Linassier C, Douziech-Eyrolles L, Marchais H, Souce M, et al. Novel method of doxorubicin-SPION reversible association for magnetic drug targeting. International Journal of Pharmaceutics 2008;363:170-176.

46. Peyre V, Spalla O, Belloni L. Compression and Reswelling of nanometric Zirconia Dispersions : Effect of Surface Complexants. J Am Ceram Soc 1997;82:1121 - 1128.

47. Kuckelhaus S, Garcia VAP, Lacava LM, Azevedo RB, Lacava ZGM, Lima ECD, et al. Biological investigation of a citrate-coated cobalt-ferrite-based magnetic fluid. J Appl Phys 2003;93:6707-6708.

48. Ojea-Jimenez I, Puntes V. Instability of Cationic Gold Nanoparticle Bioconjugates: The Role of Citrate Ions. J Am Chem Soc 2009;131:13320-13327.

49. Maiorano G, Sabella S, Sorce B, Brunetti V, Malvindi MA, Cingolani R, et al. Effects of Cell Culture Media on the Dynamic Formation of Protein-Nanoparticle Complexes and Influence on the Cellular Response. ACS Nano 2010;4:7481-7491.

50. Ballauff M, Borisov O. Polyelectrolyte brushes. Curr Opin Colloid Interface Sci 2006;11:316-323.